\documentclass[pra,showpacs,twocolumn]{revtex4}
\usepackage{dcolumn}
\usepackage{bm}
\usepackage[english]{babel}
\usepackage[babel=true]{csquotes}
\usepackage{amssymb}
\usepackage{amsmath}
\usepackage{amsfonts}
\usepackage[]{graphicx}
\usepackage{epstopdf}
\begin{document}

\title{Highly efficient singular surface plasmon generation by achiral apertures}
\author{Quanbo Jiang$^1$, Aline Pham$^1$,  Joel Bellessa$^2$, Serge Huant$^1$, Cyriaque Genet$^3$, Aur\'{e}lien Drezet$^{1\dag}$}
\address{(1)Universit\'e Grenoble Alpes, Institut NEEL, F-38000 Grenoble, France and CNRS, Institut NEEL, F-38042 Grenoble, France}
\address{(2)Institut Lumi\`{e}re Mati\`{e}re, UMR5306 Universit\'{e} Lyon 1-CNRS, Universit\'{e} de Lyon, 69622 Villeurbanne cedex, France}
\address{(3)ISIS, UMR 7006, CNRS-Universit\'e de Strasbourg, 8, all\'ee Monge, 67000 Strasbourg, France}
\email{aurelien.drezet@neel.cnrs.fr} 

\begin{abstract}
We report a highly efficient generation of singular surface plasmon (SP) field by an achiral plasmonic structure consisting of $\Lambda$-shaped apertures. Our quantitative analysis based on leakage radiation microscopy (LRM) demonstrates that the induced spin-orbit coupling can be tuned by adjusting the apex angle of the $\Lambda$-shaped aperture. Specifically, the array of $\Lambda$-shaped apertures with the apex angle $60^\circ$ is shown to give rise to the directional coupling efficiency. The ring of $\Lambda$-shaped apertures with the apex angle $60^\circ$ realized to generate the maximum extinction ratio (ER=11) for the SP singularities between two different polarization states. This result provides a more efficient way for developing SP focusing and SP vortex in the field of nanophotonics such as optical tweezers.
\end{abstract}

\maketitle

\begin{figure}[h!t]
\centering
\includegraphics[width=\columnwidth]{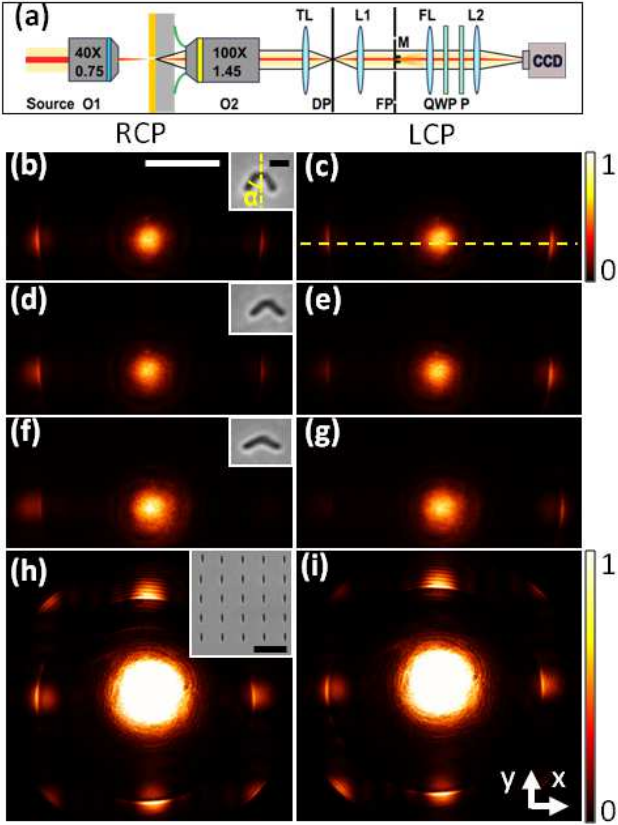}
\caption{(a) Leakage radiation microscopy setup. A normally incident laser beam is weakly focused onto the sample via a microscope objective $O_1$. The leaky SPs on direct plane (DP) or Fourier plane (FP) are collected by a CCD camera using an oil immersion objective $O_2$. The output polarization states are analysed via a polarizer (P) and quarter-wave plate (QWP). (b, g) Fourier-plane images corresponding to the excitation of arrays made of $\Lambda$-shaped apertures with $\alpha=30^\circ$ (b, c), $45^\circ$ (d, e), $60^\circ$ (f, g) as shown in the SEM images in inset (scale bar value: 200 nm). (h, i) Fourier-plane images corresponding to the excitation of arrays made of rectangular apertures. The SEM image of the whole array is displayed in inset (scale bar value: $1 \mu m$). Scale bar value in (b): $0.5k_0NA$ applied to (c-i). Color scale in (c) applied to (b-g).}    \label{image1}
\end{figure}

\indent Spin-driven singular surface plasmon generation, like surface plasmon (SP) vortices, \cite{Gor08,Yang09,Kim10,Shi12,Chen15} has been in the recent years intensively investigated in both the optical near-field \cite{Gor08,Gor09} or far-field of planar metal-dielectric nanostructures\cite{Drezet08,Gor13}. In these systems the singular generation of SPs at the metal-dielectric interface stems from the spin-orbit interactions between the incoming light and the SP modes generated. The helicity of the incident wavefront, associated with the intrinsic spin angular momentum (SAM) of photons, couples to its orbital angular momentum (OAM) via plasmonic nanostructures. SP vortices have been generated and observed in a number of planar geometries, such as  extended plasmonic Archimedes spiral, \cite{Gor08,Yang09,Chen10,Tsai14} and  chiral plasmonic nano-apertures \cite{Kim10,Tshaped,Cho12,Ku13,Ku15}.\\
\indent Among a variety of designs, the chiral T-shaped \cite{Tshaped} and the achiral $\Lambda$-shaped \cite{Chiral,Black14} antennas have attracted particular interest for their ability to support a pair of orthogonal dipoles with a phase delay. This phase shift is very sensitive to incident spin states which leads to optical spin Hall effects \cite{Bliokh08,Hosten08}. The polarization-dependent photon shift evidenced with SP propagation \cite{Gor12,Yin13} can be used for instance for inducing SP directional coupling \cite{Lee12,Rod13,Zaya14}. Recently, we used leakage radiation microscopy (LRM) on a thin metal film \cite{LRM,Hecht96,Drezet13}, in order to image singular SP vortices generated by circular structures made of T-Shaped apertures \cite{Jiang16}. This far-field methodology offers an interesting alternative to near-field measurements realized on similar systems \cite{Yin13} by allowing a precise quantitative mapping and polarization analysis of SP vortex generation in both the direct and Fourier space. In the present work, we exploit the potentials of this technique in the optimization of such systems, i.e., for controlling the contrast of the induced spin-Hall effect observed with the singular SP field generation. Specifically, we focus on $\Lambda$-shaped apertures instead of T-shaped apertures because they offer the possibility for tuning the phase delay by adjusting the apex angle $\alpha$ (we define $\alpha$ as being the half of the total angle between two elementary slits as indicated the inset in Figure \ref{image1}(b)). We firstly study the effect of varying the apex angle $\alpha$ by performing quantitative comparison between directional coupling efficiencies induced by arrays consisting of $\Lambda$-shaped apertures featuring $\alpha=30^\circ$, $45^\circ$ and $60^\circ$. Next, we confirm that this optimization procedure leads to the optimal extinction ratio of the SP singular field induced in a ring $\Lambda$-shaped apertures.

The structures investigated in this letter are milled by focused ion beam on a 50 nm thick gold film evaporated on a glass substrate. The $\Lambda$-shaped aperture consists of two rectangular slits (200 nm length and 60 nm width) separated by a distance of 150 nm for resonant excitation as explained in ref. \cite{Yin13,Jiang16}. This distance partially fixes the phase relation between two arms of the $\Lambda$-shaped aperture which in turn determines the SP directionality. By changing $\alpha$ between the two arms, we therefore tune the photon/SP coupling and the spin Hall effect strength.\\     
\indent On one hand, in order to establish the relation between the SP directionality and $\alpha$, we analyze a 5x10 array made of the aforementioned apertures. The horizontal and vertical periods of the array (Figure \ref{image1}(b,c)) are fixed at 600 nm and 300 nm for SP resonant excitation in the Ox direction and for preventing SP propagation in the Oy direction. The plasmonic arrays are illuminated by a weakly focused laser at 633 nm wavelength. The incident beam is prepared either in right circular polarization (RCP, $\sigma=-1$) or left circular polarization (LCP, $\sigma=+1$) states. The excited SPs are then recorded in the far field on a CCD (Charged Coupled Device) camera following the leakage radiation microscopy (LRM) method \cite{LRM,Hecht96,Drezet13} as sketched in Figure \ref{image1}(a).\\ \indent In order to characterize the directional coupling efficiency of the plasmonic system, we introduce a quantitative figure of merit called the directivity $V$ ($V$ is also a visibility~\cite{Jiang16}) which defines the capacity of the system to couple an incident spin into propagative SPs in a given direction. It is defined as:
\begin{equation}
V=\frac{\vert{I^\sigma_{+x}}-I^\sigma_{-x}\vert}{I^\sigma_{+x}+I^\sigma_{-x}}
 \label{eqn:visibility}
\end{equation} 
with $I^\sigma_{\pm x}$ the SP intensity launched upon $\sigma=-1$ or $+1$ incident spin and propagating towards $+x$ (right) or $-x$ (left) direction. It is recorded in the reciprocal space (see Figures \ref{image1}(b-g)) where the SP radiation appears at a well defined  leakage angle $\theta_{LR}$, hence $k$ position, corresponding to the wavevector conservation condition $n\sin{(\theta_{LR})}=\Re {\sqrt{\dfrac{\varepsilon_{metal}}{1+\varepsilon_{metal}}}}$, where $\Re$ stands for the real part, $n\simeq 1.5$ is the glass optical index, $\varepsilon_{metal}$ the metal dielectric permittivity. By performing cross sections along the center line, we determine the directivities associated with the arrays made of $30^\circ$, $45^\circ$ and $60^\circ$ apertures (the method is given in ref. \cite{Jiang16}). We measured directivities of $V_{30^\circ}=$ 0.32 $\pm$ 0.11, $V_{45^\circ}=$ 0.59 $\pm$ 0.09 and $V_{60^\circ}=$ 0.74 $\pm$ 0.07. Therefore, our findings show that the optimal apex angle maximizing the SP directivity is obtained for $\alpha=60^\circ$. Experiments with larger values of $\alpha$ (not shown) confirm that our values of $\alpha=60^\circ$ is very close to the optimum. \\
\indent In order to understand the physical mechanism at play, we analytically describe the system by means of a multi-dipolar model. The SP field resulting from the light scattering on each rectangular slit comprising a $\Lambda$-shaped aperture is regarded as induced by two pairs of in-plane dipoles oriented perpendicularly to the long and short axis of the slits. In order to quantify the contribution of each dipole, we introduced a coefficient $\beta$, defined as the relative weight of the short axis dipole amplitude with respect to the long axis. As detailed elsewhere \cite{Jiang16}, the directivity is expressed as a function of $\alpha$ and $\beta$ as:
\begin{equation}
V=\frac{2\beta(1-\beta)\tan^3\alpha+2(1-\beta)\tan\alpha}{\beta^2\tan^4\alpha+(1+\beta^2)\tan^2\alpha+1}
 \label{eqn:visibilitybeta}
\end{equation}

Experimentally, the coefficient $\beta$ can be determined using a reference array made of vertically oriented ($\alpha=0^\circ$) rectangular slits with dimensions similar as that of the $\Lambda$-shaped aperture's arms (Figure \ref{image1}(f)). The horizontal and vertical periods are fixed at 600 nm so as to generate propagative SP in both Ox and Oy directions. The ratio of the SP intensities along the two orthogonal directions in the Fourier plane directly leads to $|\beta|^2$, i.e. the ratio of the two dipolar contributions. A value $\beta=0.50\pm0.07$ is measured allowing us to derive a theoretical value of $\alpha$ for which the directivity is maximum. In agreement with the experimental results, we find that the angle $\alpha=60^\circ$ yields the maximum value of the directivity. Under the single dipole approximation ($\beta=0$), we retrieve the results obtained by previous numerical simulations \cite{Huang15} which predicted a maximum directivity for apertures with $\alpha=45^\circ$. 

\begin{figure}[h!t]
\centering
\includegraphics[width=\columnwidth]{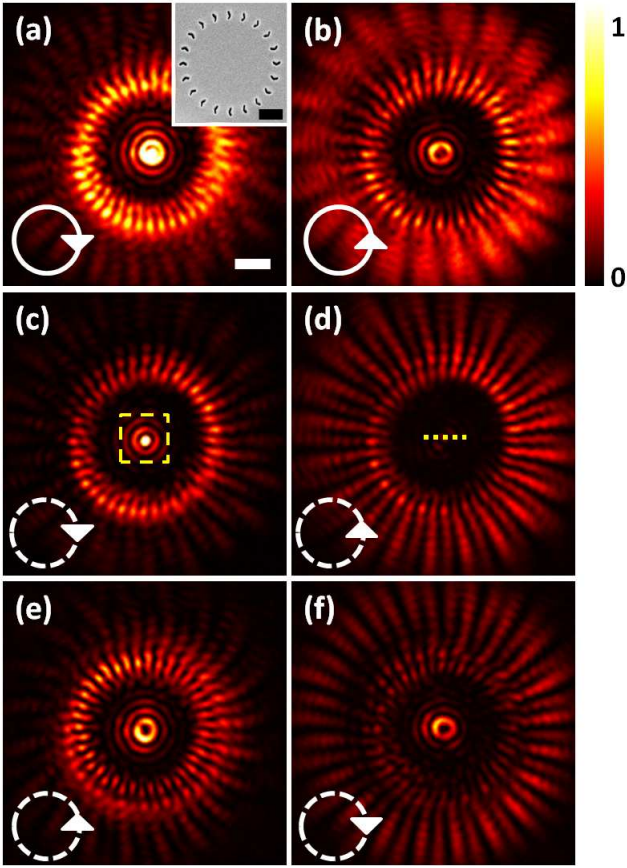}
\caption{Direct-plane images for the circle of $\Lambda$-shaped apertures with $\alpha=60^\circ$. A beam block in the Fourier plane is used to remove the directly transmitted light from the incident beam. Right (respectively left) column (a, b) Signal recorded under the polarization excitation indicated by the solid arrows. (c, e) Signal recorded under RCP with polarization analysis in the circular basis as indicated by the dotted arrows. (d, f) Signal recorded under LCP with polarization analysis. The inset in (a) is the SEM image for the circle of $\Lambda$-shaped apertures. The scale bars in (a) are both 1 $\mu m$ }    \label{image2}
\end{figure}

We will now demonstrate that our method can be implemented in the design of plasmonic structures in order to improve singular field generation efficiency. Indeed, it has been reported that chiral T-shaped apertures arranged in a circular geometry induce inward or outward radial SP coupling according to the spin of the excitation beam \cite{Tshaped,Jiang16}. Here, we employ achiral $\Lambda$-shaped apertures instead to reach the maximum efficiency by optimizing the apex angle (Figure \ref{image2}(a)). Also, since the symmetry of the $\Lambda$-shaped aperture in the array is simpler than that for the T-shaped one the optimization will be easier to implement. We emphasize that the structure is globally chiral since the circle is oriented by the direction of the $\Lambda$-shaped apertures.  We first fabricate a plasmonic structure formed by 20 rotating $\Lambda$-shaped apertures with the apex angle $\alpha=60^\circ$. The latter is illuminated with a circular polarized beam and the SP field is recorded in the image plane. In order to selectively collect the SP signal, a spatial filter is placed in the Fourier plane. On Fig \ref{image2} (a), (b), a spin-sensitive response of the SP clearly appears upon RCP and LCP excitation as a result of spin-orbit coupling. In particular, a bright (dark) central spot is observed when the signal features the same (opposite) handedness as the input field (Fig \ref{image2} (c), (d)). We emphasize that the SP focusing and vortices are more intense under RCP than LCP excitation because the right-handed structure of $\Lambda$-shaped apertures inwardly generates SPP under RCP excitation. These recorded intensity distributions can be understood by applying the above analytical study to a ring of $\Lambda$-shaped apertures. Specifically, we show that the singular SP intensity near the origin upon an illumination with the spin state $\sigma$ is a combination of Bessel's functions \cite{Jiang16} such as $I_\sigma^{tot}=I_{\sigma,L}+I_{\sigma,R}$ with:

\begin{equation}
\begin{split}
I_{\sigma,L}\propto \mid C_{\sigma}\mid ^2 J_{\sigma-1}^2(k_{SP},\rho )\\
I_{\sigma,R}\propto \mid C_{\sigma}\mid ^2 J_{\sigma+1}^2(k_{SP},\rho )\\
\end{split}
\end{equation}

$I_{\sigma,j}$ denotes the resultant SP intensity launched by LCP/RCP polarization states ($\sigma=+1/-1$) and analyzed in RCP/LCP states ($j=L/R$). $\rho$ represents the distance between the $\Lambda$-shaped elements and the circular structure center, and $J_{\sigma\pm1}$ stands for the $(\sigma\pm1)^{th}$ order Bessel function \cite{Kim10,Yang13}. In the case of $\sigma=\pm1$, $J_0^2$ and $J_{\pm2}^2$ represent that the output polarization analysis is similar and opposite to the incident polarization states respectively. $C_{\sigma}$ ($\sigma=\pm1$) refers to the SP coupling efficiency of the single $\Lambda$-shaped element upon RCP (LCP) polarization input state given by \cite{Jiang16}:
\begin{equation}
\begin{aligned}
C_{\sigma}=1-\sigma\tan\alpha+(\tan\alpha-\sigma)i\\
+\beta[1+\sigma{\tan\alpha}-(\tan\alpha+\sigma)i]
\end{aligned}
\label{eqn:efficiency}
\end{equation}

\begin{figure}[h!t]
\centering
\includegraphics[width=\columnwidth]{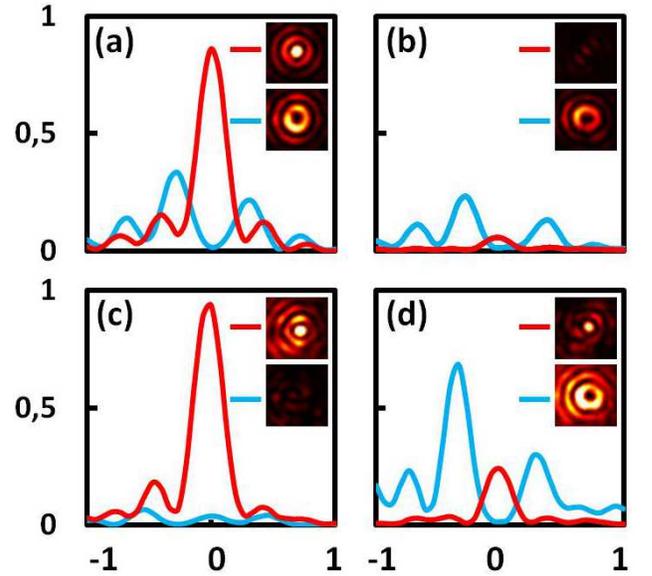}
\caption{Intensity cross-section along center-line as indicated by the yellow dashed line in Figure \ref{image2}(d). (a, b) Cross-section polarized analysis for $\alpha=60^\circ$ and (c, d) cross-section polarised analysis for $\alpha=30^\circ$. Zoom on SP focusing spot and SP vortex are shown in insets as indicated by the yellow dotted box in Figure \ref{image2}(c). (a, c) are the results under RCP input state and (b, d) are under LCP input state. The red curves correspond to the same output state (polarizer) with the input state. The blue curves correspond to the reversed output state (polarizer) with the input state.}    \label{image3}
\end{figure}

The extinction ratio defined as $ER=\mid C_{-1}\mid ^2/\mid C_{+1}\mid ^2$ is used to quantify the singular SP generation induced by our plasmonics structure. Theoritically, for the experimental value of $\beta$=0.5, the apex angle that optimizes $ER$ is predicted for $\alpha=60^\circ$ with $ER_{60^\circ}=13.93$. In order to experimentally determine $ER$, we derive the ratio between the central peak values from cross-sections intensities as depicted in Figure \ref{image3}(a),(b). In consistence with the theory, we find $ER_{60^\circ}=11.02\pm3.22$ which is higher than the extinction ratio of the ring of T-shaped apertures previously reported ER=7.45 \cite{Jiang16}. It demonstrates that the circular ring design of achiral $\Lambda$-shaped apertures with the optimal apex angle has the same effect on singular SP generation and even better efficiency than the structure of chiral T-shaped apertures. Furthermore, we verify that this design corresponds to the optimal design by comparing it to structures fabricated with the apex angle $\alpha=30^\circ$ and $45^\circ$. The later are shown to feature $ER_{30^\circ}=3.90\pm2.17$ and $ER_{45^\circ}=6.45\pm3.01$, which is in agreement with the expected theoretical values $ER_{30^\circ}=2.18$ and $ER_{45^\circ}=4.00$ as well as with the directional coupling efficiency of the arrays. As predicted, the SP vortices present two side lobes with a minimum intensity or singularity at the center as indicated by the blue curves in Figure \ref{image3}. Noteworthy, our model does not take into account the coupling between adjacent apertures and the SP field around the center approximates at the origin, which could explain the observed discrepancies between the experimental and theoretical data.\\ 
\indent In summary, the optimal apex angle of the achiral $\Lambda$-shaped structure was successfully determined for both directional coupling and singular SP generation in the far field. Our method based on LRM detection allows quantitative analysis and was proven to be a sophisticated characterization technique for mapping SP vortex field. It provides several new possibilities for polarization controlled SP subwavelength focusing. The presented multi-dipolar model was demonstrated to be a suitable tool for predicting the extinction ratio of the SP singularity. In particular, it highlights the the short axis aperture contribution has to be taken into account in the design of directional plasmonic structure. All these findings offer a promising way for device development in the field of nanophotonics such as optical tweezers,\cite{Rig07,Pang12} particle trapping \cite{Liu10,Wang11} etc.

\section*{Funding Information}
\indent This work was supported by Agence Nationale de la Recherche (ANR), France, through the SINPHONIE (ANR-12-NANO-0019) and
PLACORE (ANR-13-BS10-0007) grants. The Ph.D.
grants of Q. Jiang by the R\'egion Rh\^one-Alpes and of A. Pham by the Minist\`ere de l'enseignement et la recherche, scientifique, are gratefully
acknowledged.
\section*{Acknowledgments}
 We thank J.-F. Motte and G.~Julie, from NANOFAB facility in Neel Institute, for sample fabrication.




\end{document}